\documentclass[aps,prl,twocolumn,superscriptaddress,showpacs]{revtex4}
\usepackage{amsmath}
\usepackage{amssymb}
\usepackage{graphicx}
\usepackage{ulem}

\begin{document}
\title{Triplet pair correlations in {\it s-}wave superfluids as a signature of the FFLO state}
\author{I. Zapata}
\affiliation{Departamento de F\'{\i}sica de Materiales, Universidad Complutense de Madrid,
E-28040 Madrid, Spain}
\affiliation{Department of Physics, Harvard University, 17 Oxford Street, Cambridge, MA 02138, USA}
\author{F. Sols}
\affiliation{Departamento de F\'{\i}sica de Materiales, Universidad Complutense de Madrid,
E-28040 Madrid, Spain}
\author{E. Demler}
\affiliation{Department of Physics, Harvard University, 17 Oxford Street, Cambridge, MA 02138, USA}

\pacs{67.85.-d,03.75.Ss,71.10.Pm,74.45.+c}
\date{\today}

\begin{abstract}
We show that anti-parallel triplet pairing correlations are generated in superfluids with purely {\it s-}wave interactions whenever population imbalance enforces anisotropic Fulde-Ferrell (FF) or inhomogeneous Larkin-Ovchinikov (LO) states. These triplet correlations appear in the Cooper pair wave function, while the triplet part of the gap remains zero. The same set of quasiparticle states contributes to the triplet component and to the polarization, thus spatially correlating them. In the LO case, this set forms a narrow band of Andreev states centered on the nodes of the {\it s-}wave order parameter. This picture naturally provides a unifying explanation of previous findings that attractive {\it p}-wave interaction stabilizes FFLO states. We also study a similar triplet mixing which occurs when a balanced two-component system displays FFLO type oscillations due to a spin-dependent optical lattice. We discuss how this triplet component can be measured in systems of ultra-cold atoms using a rapid ramp across a {\it p-}wave Feshbach resonance. This should provide a smoking gun signature of FFLO states.
\end{abstract}
\volumeyear{2011}
\volumenumber{number}
\issuenumber{number}
\eid{identifier}
\startpage{1}
\endpage{ }
\maketitle

Conventional (homogeneous and unpolarized) BCS states survive in the presence of a Zeeman field as long as the condensation energy is larger than polarization energy gain in the normal phase \cite{clogston1962}. A deep argumentation of why these two are the only allowed homogenous and isotropic configurations for usual {\it s-}wave superconductors can be found in Ref. \cite{leggett2006}. Fulde-Ferrel (FF) \cite{fulde1964} and Larkin-Ovchinnikov (LO) \cite{larkin1965} found that a compromise between BCS condensation and polarization could occur in a region of the phase space lying between the unpolarized BCS and the polarized normal Fermi gas phases. Those states are characterized by the Cooper pair wave function having one (FF) or two (LO) Fourier components of nonzero total momentum. As a result, rotational (FF and LO) and translational (LO) symmetries are spontaneously broken.

The FFLO state is expected to occur in a variety of contexts including heavy fermion systems \cite{matsuda2007}, high density QCD \cite{rajagopal2000} and neutron stars \cite{casalbuoni2004}. However, there is much controversy on the existence of FFLO states in superconducting systems \cite{buzdin2005}. Thus, there is a strong motivation to explore this intriguing state of matter in systems that allow detailed control of their parameters, such as ultracold fermions \cite{radzihovsky2010}. Moreover, cold atomic gases can be used to implement experimental probes that are very different from what is available in electron systems. Hence, experiments with ultracold atoms can provide unique signatures of exotic many-body states. The goal of this paper is to show that an unequivocal signature of the FFLO phase can be obtained through direct measurements of the triplet pairing component in the FFLO phase in cold atomic gases.
In this paper we show that triplet correlations always accompany FFLO-type states. These correlations are very difficult to probe in electron systems (see however \cite{eschrig2011} and references therein). On the other hand, we show how a variation of the technique of rapidly ramping along a (in our case, {\it p}-wave) Feshbach resonance gives a direct access to these triplet correlations. This method can be used for an independent test of FFLO physics beyond those already proposed \cite{edge2009,loh2010,edge2010,lutchyn2010}. Our method can be implemented with current experimental techniques \cite{liao2010,huletPC}.

Another motivation for our analysis comes from recent theoretical studies, which found that {\it p}-wave attractive interactions are surprisingly efficient in stabilizing the FFLO state. Even when the {\it p}-wave interaction is but a small fraction of the dominant {\it s-}wave attraction, it considerably enhances triplet pair correlations at temperatures well above the nominal {\it p}-wave attraction critical temperature. Such behavior appears in low-dimensional superconductors under a parallel magnetic field (which yields a Zeeman but not an orbital effect) \cite{matsuo1994,shimahara2000,shimahara2002} and dipolar cold gases \cite{samokhin2006}. Triplet correlations in the presence of attractive {\it p}-wave interactions seem to reside in a narrow band of Andreev states, which also host the polarization and which are localized near the nodes of the {\it s-}wave gap \cite{burkhardt1994,zheng2010}.

Earlier analysis \cite{matsuo1994,shimahara2000,shimahara2002,samokhin2006,zheng2010} was based on the assumption that the presence of an attractive {\it p}-wave channel is necessary to find triplet pair correlations (see however \cite{dutta2008} for the case of Abrikosov lattices). In the present work we argue that a discussion centered on the gap tends to miss an important property of the Cooper pairs, namely, that triplet correlations are independently produced, even without the help of interactions in the {\it p}-wave channel. In fact, we show that an inhomogeneous {\it s-}wave order parameter together with polarization is sufficient to generate a triplet component in the Cooper pair. The effect of any residual {\it p}-wave interaction is to lower the energy of the FFLO state using the already macroscopic occupation of the triplet component.

We consider two ways of achieving FFLO type states, spin imbalance and spin-dependent optical lattices \cite{zapata2010}. For simplicity of presentation we focus on 1D scenarios of neutral two-species Fermi superfluids that show a competition between traditional {\it s-}wave pairing and paramagnetic order without the intervention of the magnetic orbital (Meissner) effect. Importantly, the interaction between species reduces to an {\it s-}wave (singlet) channel, namely, $g_{\rm 1D}\int dx \Psi_{\uparrow}^{\dag}(x)\Psi_{\downarrow}^{\dag}(x)\Psi_{\downarrow}(x)\Psi_{\uparrow}(x)$, where $g_{\rm 1D}$ is the effective 1D coupling constant. We will use $g_{\rm 1D}k_F/E_F=-2.04$ as in Refs. \cite{drummond2007, zapata2010}, corresponding to an
interaction to kinetic energy ratio of $-m g_{\rm 1D}/\hbar^2=1.6$.

It has been recently noted \cite{bulgac2006,gaudio2007,bulgac2009,patton2011} that, in the normal (non-BCS condensed) phase, the intra-species interaction
can be strongly modified (and in particular turned attractive) by the polarization of the medium. As a consequence, at sufficiently low temperatures, normal phases are unstable against the formation of intra-species {\it p}-wave superfluids. In the unpolarized case considered in Ref. \cite{gaudio2007}, the {\it s}-wave interaction always dominates over the effectively induced parallel-spin {\it p}-wave attraction. We conjecture that in such a case, once the {\it s}-wave superfluid is formed, the resulting reduction of the spin fluctuations (due to the formed Cooper pairs) will lower the effective parallel spin {\it p}-wave attraction more than envisaged in Ref. \cite{gaudio2007}. Thus we have chosen to work with systems where the induced {\it p}-wave attraction can be neglected.

To develop a microscopic description of the system, we introduce the Bogoliubov transformation $\Psi_{\sigma}(x,t)=\sum_{k}[u_{k \sigma}(x)e^{-i \omega_{k \sigma} t} c_{k \sigma}+\sigma
v^{*}_{k \bar{\sigma}}(x)e^{i \omega_{k \bar{\sigma}} t} c_{k \bar{\sigma}}^{\dag}]$  where $\bar{\sigma}:=-\sigma$, $c$ and $c^\dag$ denote the quasiparticles and the sum should be restricted to avoid double counting (we use $\omega_{k \sigma}>0$ with $k$ the quasiparticle index composed of a quasimomentum and the band index), which yields the Bogoliubov - de Gennes (BdG) equations for the quasiparticle wave functions:
\begin{eqnarray}\label{eqnBdG}
\nonumber
  &&\left[
      \begin{array}{cc}
        H_{\sigma} & \Delta(x) \\
        \Delta^*(x) & -H_{\bar{\sigma}} \\
      \end{array}
    \right]
    \left[
      \begin{array}{c}
        u_{k \sigma}(x) \\
        v_{k \sigma}(x) \\
      \end{array}
    \right] = \omega_{k \sigma}
    \left[
      \begin{array}{c}
        u_{k \sigma}(x) \\
        v_{k \sigma}(x) \\
      \end{array}
    \right], \\
  &&H_{\sigma} = -\frac{\hbar^2}{2m}\frac{\partial^2}{\partial x^2}+V_{\sigma}(x)-\mu_{\sigma}+g_{\rm 1D}n_{\bar{\sigma}}(x),
\end{eqnarray}
where we have introduced the one-species density, $n_{\sigma}(x):=n_{\sigma}(x,x)$, with $n_{\sigma}(x,x'):=\langle\Psi_{\sigma}^{\dag}(x)\Psi_{\sigma}(x')\rangle$ the one-particle density matrix (normal average), and the {\it s-}wave gap, $\Delta(x):=-g_{\rm 1D}F(x,x)$, with $F(x,x') := \langle\Psi_{\downarrow}(x)\Psi_{\uparrow}(x')\rangle$ the pair wave function (anomalous average) \cite{note1}. These normal and anomalous averages read
\begin{eqnarray}\label{eqnBdGSC}
\nonumber
n_{\uparrow}(x,x')&=&\sum_{\omega_{k \uparrow}} f(\omega_{k \uparrow}) u^*_{k \uparrow}(x) u_{k \uparrow}(x') \\
\nonumber
n_{\downarrow}(x,x')&=&\sum_{\omega_{k \uparrow}} f(-\omega_{k \uparrow}) v_{k \uparrow}(x) v^*_{k \uparrow}(x') \\
F(x,x')&=&-\sum_{\omega_{k \uparrow}} f(\omega_{k \uparrow}) v^*_{k \uparrow}(x) u_{k \uparrow}(x'),
\end{eqnarray}
where $f(\omega_{k \sigma})\equiv [1+\exp(\hbar \omega_{k \sigma}/k_B T)]^{-1}$, and are to be solved self-consistently. Notice that these expressions only contain $u_{k\sigma},v_{k\sigma},\omega_{k\sigma}$ for $\sigma=\uparrow$ and the sums are unrestricted \cite{drummond2007}.

It can be shown (see Ref.\cite{leggett2006}) that the only macroscopically occupied eigenstate $\phi_{\alpha\beta}(x_1,x_2)$ of the two-particle density matrix is trivially related to the pair wave function via a scaling $F(x_1,x_2)=\sqrt{N_0}\phi_{\downarrow\uparrow}(x_1,x_2)$, where $\phi_{\alpha\bar{\alpha}}(x,x')=-\phi_{\bar{\alpha}\alpha}(x',x)$ is the wave function of the Cooper pairs, normalized as $\sum_{\alpha\beta}\int dx_1 dx_2 |\phi_{\alpha\beta}(x_1,x_2)|^2=1$, and $N_0$ is the number of Cooper pairs. Its singlet/triplet components are given by $F^{s/t}(x_1,x_2)=-[F(x_1,x_2)\pm F(x_2,x_1)]/2$. Shifting to the Cooper pair center-of-mass and relative position coordinates, $X:=(x_1+x_2)/2, y:=x_1-x_2, F^{s/t}(X;y):=F^{s/t}(X+y/2,X-y/2)$, and using fermionic permutation symmetry, it is immediate to see that $F^t(X;0)=0$ and $\partial F^s(X;y)/\partial y|_{y=0}=0$. The quantity $F^s(X;0)=\Delta(X)/g_{\rm 1D}$ is directly related to the conversion into {\it s-}wave molecules when ramping through an {\it s-}wave Feshbach resonance (see \cite{zapata2010}). In this paper we will demonstrate that $G^t(X):=\partial F^t(X;y)/\partial y|_{y=0}$ measures the conversion to {\it p}-wave molecules when ramping through {\it p}-wave Feshbach resonances (see below).

We start by considering spin imbalanced systems without spin-dependent potentials, $V_\sigma(x)=0$.
A simple explanation of the emergence of triplet pair correlations can be attained by studying Eq. (\ref{eqnBdG}), for $\sigma=\uparrow$, in the particular case where the Hartree terms are ignored, with the chemical potential difference playing the role of an effective exchange (Zeeman) field $h:=\mu_{\uparrow}-\mu_{\downarrow}$. As a result, $H_\sigma \rightarrow H+\sigma h/2$, where $H=-(\hbar^2/2m)(\partial^2/\partial x^2)-\mu$ with $\mu \equiv (\mu_{\uparrow}+\mu_{\downarrow})/2$. The resulting eigenvalues are $\omega_{k\uparrow}=\epsilon_{k}+ h/2$, with $\epsilon_{k}$ the eigenvalue for zero exchange field. This problem shows a further property which is common to unpolarized systems \cite{degennes1999}: given a $\epsilon_k$ solution, $\chi_k(x):=(\left(u_k(x), v_k(x)\right)^\top$, there is another one with eigenvalue $-\epsilon_k$, and wave function $-i\sigma_y \chi_k^*(x)=\left(-v^*_k(x), u^*_k(x)\right)^\top$. The identity can then be written:
\begin{equation}
\delta(x-x')=\sum_{\epsilon_k>0} \left(
\chi_k(x) \chi_k^\dagger(x') + \sigma_y\chi_k^*(x) \chi_k^\top(x') \sigma_y \right)\;.
\label{eqnBdGComplet}
\end{equation}
After some algebra, we find for the anomalous singlet/triplet pairing components and polarization at zero-temperature:
\begin{eqnarray}
\nonumber
F^s(x,x')&=&-\frac{1}{2} \sum_{\epsilon_k>h/2} [v^*_k(x')u_k(x) + v^*_k(x)u_k(x')] \\ \nonumber
F^t(x,x')&=& -\frac{1}{2} \sum_{0<\epsilon_k<h/2} [v^*_k(x')u_k(x) - v^*_k(x)u_k(x')] \\
p(x)&=& -\sum_{0<\epsilon_k<h/2} [|u_k(x)|^2 + |v_k(x)|^2].
\label{eqnTripletSingletPolPWF}
\end{eqnarray}
where $p(x)\equiv n_{\uparrow}(x)-n_{\downarrow}(x)$. Equation (\ref{eqnTripletSingletPolPWF}) clearly reveals that triplet correlations and spin polarization are closely related because they are supported by the same set of quasiparticle states. It is known \cite{burkhardt1994,vorontsov2005} that Andreev sub-gap states exist (in the LO case) around the nodes of the gap. According to the present argument, they contribute to polarization and triplet correlations only when a Zeeman field is present.

\begin{figure}[htb!]
\includegraphics[width=\columnwidth]{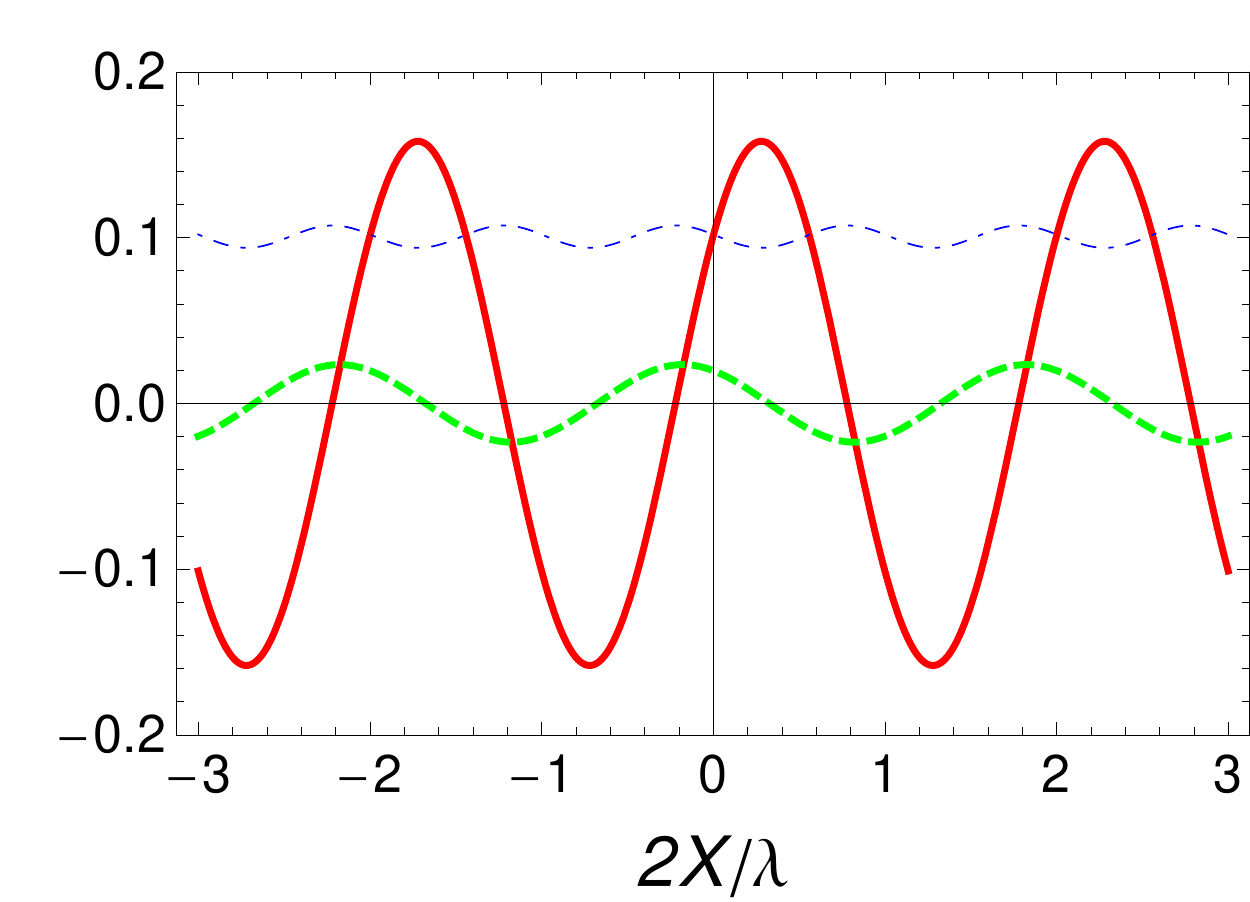}
\caption{The polarization $(n_{\uparrow}(X)-n_{\downarrow}(X))/k_F$ in blue dashed-dotted, pairing $\Delta (X)/E_F$ in red and triplet order parameter $G^t(X)/k_F^2$ in green, as functions of $2X/\lambda$ for a LO state with a global polarization $(n_{\uparrow}-n_{\downarrow})=0.16 2k_F/\pi$, $k_F \lambda=20, T=0$ and  $g_{\rm 1D}k_F/E_F=-2.04$.}
\label{figLOTriplet}
\end{figure}

A fully self-consistent computation of a LO state is shown in Fig. \ref{figLOTriplet}. The triplet-condensate fraction for this case is $\int  dX dy |F^t(X;y)|^2/N_0=0.42$. A clear correlation is observed between the maxima of the polarization $p(x)$ and those of the triplet correlation profile $G^t(x)$, both peaking near the nodes of the {\it s-}wave gap.

\begin{figure}[htb!]
\includegraphics[width=\columnwidth]{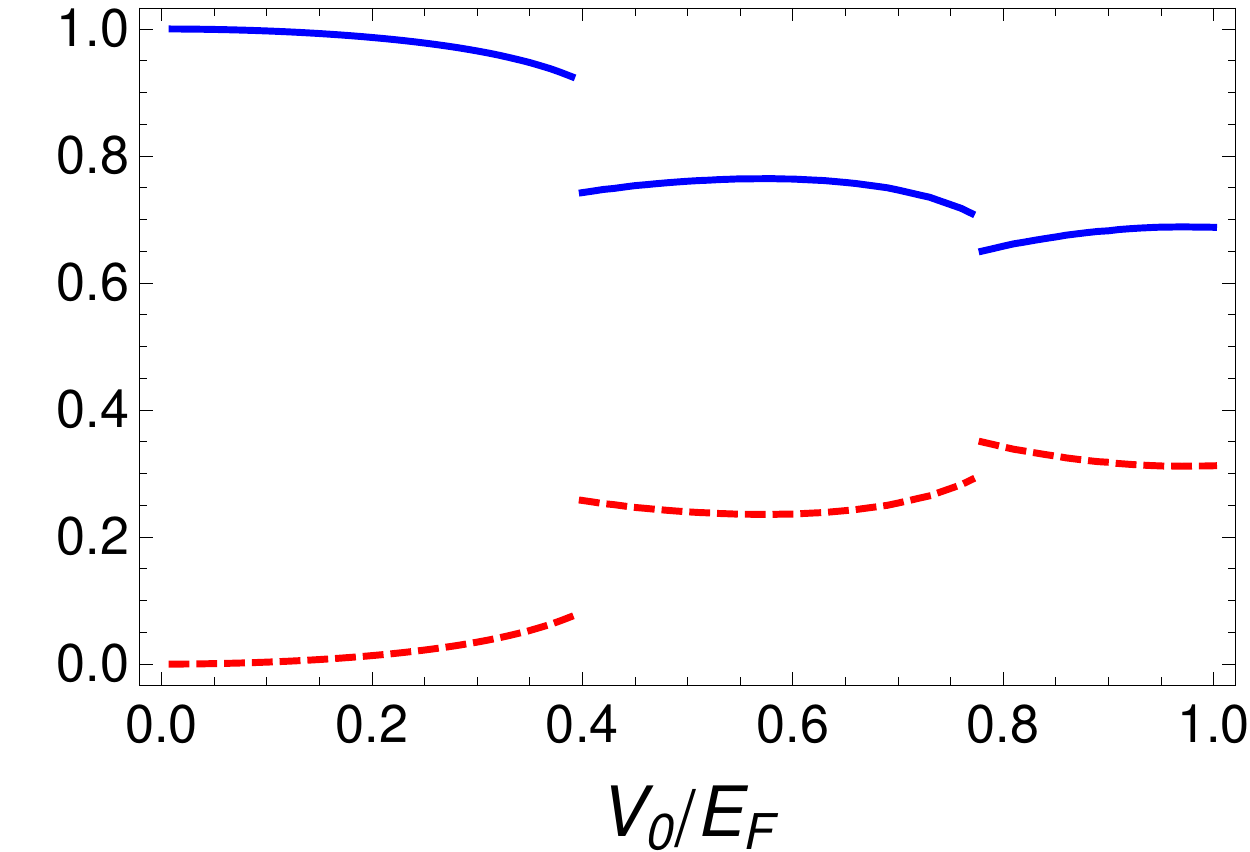}
\caption{Singlet ($\int dy dX |F^s(X;y)|^2/N_0$, blue) vs triplet ($\int dy dX |F^t(X;y)|^2/N_0$, red) condensate fractions for a $\pi$-phases structure in a spin-dependent lattice potential of wavelength $\lambda=30$ as a function of the optical lattice strength $V_0/E_F$, for interaction strength $g_{\rm 1D}k_F/E_F=-2.04$ and zero temperature. }
\label{figTripletSingleFractions}
\end{figure}

 We now discuss a different route to FFLO type states based on a spin-dependent optical potential \cite{zapata2010} $V_{\uparrow}(x)=-V_{\downarrow}(x)= V_0 \cos(2\pi x/\lambda)$. Cooper pairs experience stretching forces which peak at the zeros of the potential, i.e. each spin species is attracted to its respective minimum. The resulting elongation of the pairs generates local triplet {\it p}-wave mixing and a lowering of the gap amplitude. As $V_0$ increases, $\pi$-phases (zeros of the gap) appear at some critical values of the potential amplitude. Then the triplet weight is displaced and centered around the new gap nodes, changing the configuration abruptly. This results in a discontinuous behavior of the total triplet and singlet pair components, as can be seen in Fig. \ref{figTripletSingleFractions}. Successive jumps take place when a transition to a higher order $\pi$-phase occurs. Clearly, the zeros of the gap favor the appearance of triplet correlations.


\begin{figure}[htb!]
\includegraphics[width=\columnwidth]{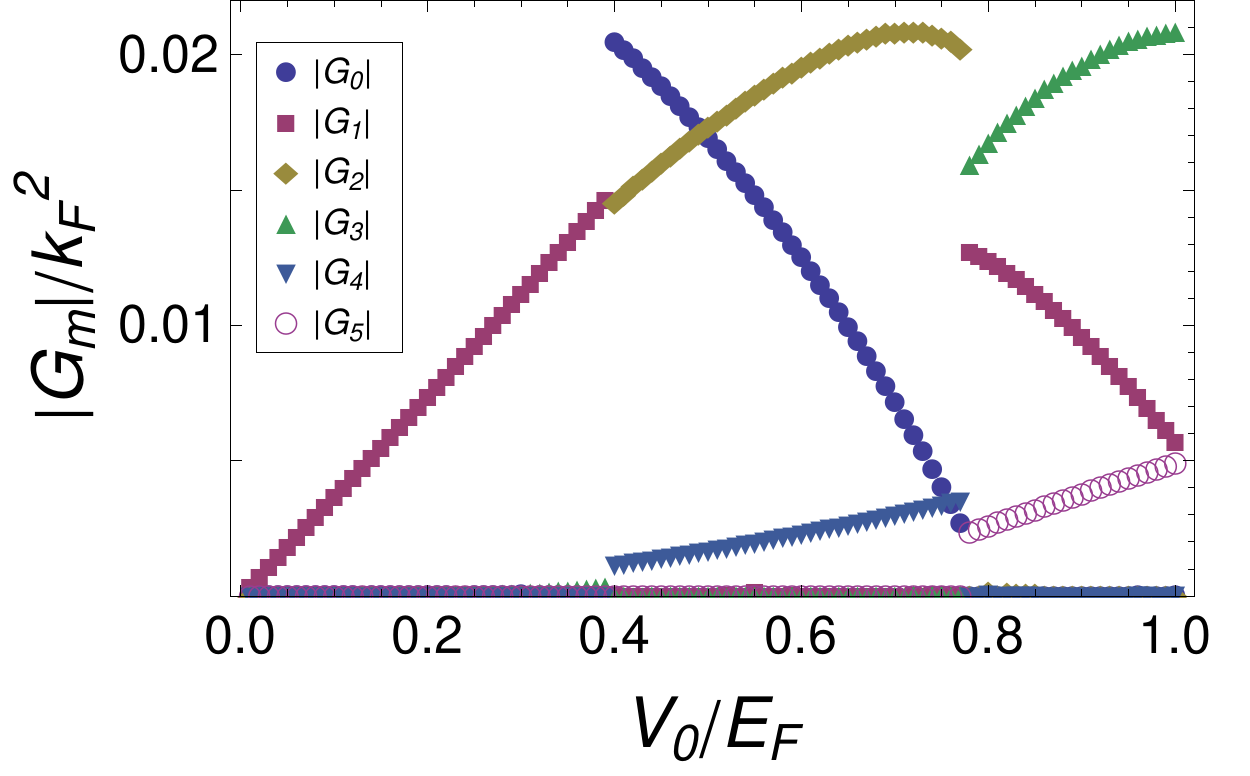}
\caption{$\pi$-phases setup plot showing the absolute value of the Fourier components, $G_m:= G_{q_x=2\pi m/\lambda}$ from Eq.~\eqref{eqnMolNumbFinal}, of the triplet order parameter, as function of the lattice potential strength $V_0$ (parameters as in Fig.~~\ref{figTripletSingleFractions}). Only odd components are non-zero for $V_0/E_F\leq0.39$, whereas only even ones are non-zero for $0.39\leq V_0/E_F\leq 0.69$, and so forth (opposite behavior as the gap, see \cite{zapata2010}).}
\label{figAbsDF}
\end{figure}

In Fig. \ref{figAbsDF} we show the absolute value of the Fourier components of the triplet order parameter. These quantities are important because, as we show below, they directly correlate with the number of {\it p}-wave molecules obtained after a ramp. Specifically, to observe the triplet correlations predicted in the present work, we propose a combination of a $\pi/2$ RF-pulse followed by a ramp over a {\it p}-wave Feshbach resonance, which is a variation of a well known technique \cite{greiner2003,regal2004,zwierlein2004,diener2004,altman2005}.

In general, {\it p}-wave Feshbach resonances can be found both between identical or between different hyperfine species. However, in the case of $^{40}$K \cite{chin2010} there is no interspecies {\it p}-wave Feshbach resonances yet found. $^6$Li does have those resonances, but very far away from the 834 Gauss {\it s-}wave resonance (or from the other {\it s-}wave Feshbach resonances). However $^{40}$K shows a double {\it p}-wave Feshbach resonance among bb hyperfine states at 198.4 and 198.8 Gauss, very close to the 202.1 Gauss {\it s-}wave Feshbach resonance. We might use a $\pi/2$ RF-pulse to partially convert some of the Cooper pairs (made of different hyperfine species) into homo-species pairs and, before any relaxation takes place, ramp over a {\it p}-wave Feshbach resonance (between same hyperfine species) to convert them into {\it p}-wave molecules.

To estimate the number of {\it p}-wave molecules (of bound energy $\epsilon=-\hbar^2/m a_p^2$ and center of mass momentum $\hbar \boldsymbol{q}$) formed it is enough to compute the average of the number of molecules operator in the state before the ramp (followed by the $\pi/2$-pulse, $\hat{U}_{\pi/2}^\dag \Psi_\uparrow \hat{U}_{\pi/2}=(\Psi_\uparrow+\Psi_\downarrow)/\sqrt{2}$, if needed) \cite{diener2004,altman2005}:
\begin{eqnarray}\label{eqnMolecCreation}
\nonumber
n_{\boldsymbol{q}\epsilon}&:=&\sum_{m=-1,0,1}\langle b^\dag_{\boldsymbol{q}\epsilon m} b_{\boldsymbol{q}\epsilon m}\rangle \\ \nonumber
b^\dag_{\boldsymbol{q}\epsilon m}&:=&\sum_{\alpha, \beta=\uparrow,\downarrow} \int d^3r_1 d^3r_2 \frac{e^{i \boldsymbol{q}\cdot(\boldsymbol{r}_1+\boldsymbol{r}_2)/2}}{\sqrt{L^3}} \phi_{\epsilon m}(\boldsymbol{r}_1-\boldsymbol{r}_2)_{\alpha \beta} \\
&&\Psi_\alpha^\dag (\boldsymbol{r}_1) \Psi_\beta^\dag (\boldsymbol{r}_2),
\end{eqnarray}
here $\phi_{\epsilon m}(\boldsymbol{r})_{\alpha\beta}$ denotes the spinor $\alpha, \beta$ components of the molecular orbital {\it p-}wave function of angular momentum projection quantum number $m=-1,0,1$ and energy $\epsilon$ and $L^3$ is the volume. Assuming quasi-one dimensional configuration $\Psi^\dag_\sigma(x)=\int d^2 \rho \phi(\boldsymbol{\rho})\Psi^\dag_\sigma(x, \boldsymbol{\rho})$, where $\phi(\boldsymbol{\rho})=\exp[-\rho^2/(2 a_\bot^2)]/(\sqrt{\pi} a_\bot)$ is the transverse confinement wave function and $\boldsymbol{\rho}$ refers to the transverse coordinates, $\boldsymbol{r}=(x, \boldsymbol{\rho})$, the averages can be factorized as $\langle \Psi_{\downarrow}(\boldsymbol{r}_1) \Psi_{\uparrow}(\boldsymbol{r}_2)\rangle=\phi(\boldsymbol{\rho}_1) \phi(\boldsymbol{\rho}_2) F(x_1,x_2)$, the same occurring to the one-particle density matrix terms. Surviving terms from the one-particle density matrix $n_{\sigma}(x_1,x_2)$ can be shown to lead to a structureless continuous background. The reason is that they always contain one coordinate from each molecular state involved. The other terms, which come from the pair wave function before the RF pulse, conspire to depend on its triplet part alone and each depends on one set of molecular state variables ({\it s-}wave components of the Cooper pair are lost due to contraction with the {\it p-}wave symmetry molecular state). The length scales of variation of $F(X;y)$ as a function of $y$ are $k_F^{-1}$ and the BCS coherence length $\xi=\hbar v_F/\Delta_0$, with $\Delta_0$ the zero temperature gap. We assume that the the other scales are much shorter, namely $k_F a_\bot, k_F a_p \ll 1, k_F \xi$.  Then, we can Taylor expand $F^t(X;y)$ in the $y$ variable and take the lowest non-zero term, which by symmetry is the gradient $F^t(X;y)\simeq y  G^t(X)$.  The final result is:
\begin{eqnarray}\label{eqnMolNumbFinal}
\nonumber
n_{q_x}&=&L \left| G_{q_x} \right|^2 16 \gamma a_p a_\bot^2 \\
G_{q_x}&:=&\frac{1}{L}\int dX e^{i q_x X} G^t(X),
\end{eqnarray}
where $\gamma$ is a dimensionless factor which can be further estimated as $\gamma\simeq3$ for $a_p\gg a_\bot$, and $\gamma\simeq (a_p/a_\bot)^4/3$ for $a_p \ll a_\bot$. In the case in which the {\it p}-wave Feshbach resonance allows for interspecies molecular formation, there is no need of $\pi/2$ pulse and the results are the same but with a signal four times bigger.

In the case \cite{moritz2005} of $^{40}$K, with a 1D geometry of tubes that are optically trapped with a superposed magnetic field to control the interaction via the Feshbach resonance at $B_0=202.1$G and $N\sim 100$ per tube of length $L\sim 40\mu\text{m}$, we have $n\sim2.5$ $\mu$m$^{-1}$ and $k_F=\pi n/2\sim 3.93$ $\mu$m$^{-1}$. For simplicity
we neglect the external confinement. Using $a_\bot \sim 60.3$nm, $a_p\sim132.3$nm and $g_{1D}kF/EF=-2.04$ we find $n_{q_x=2\pi/\lambda}\sim0.09$ for $G_{q_x}\sim 0.02 k_F^2$ with momentum $q_x\sim 1.23\mu$m, like in Fig. \ref{figLOTriplet}. However, the number of tubes in \cite{moritz2005} was 4900 so we expect a signal of 440 molecules, which should be detectable.

The effect of induced parallel-spin {\it p}-wave interactions \cite{bulgac2006,gaudio2007,bulgac2009,patton2011} on the symmetry of the Cooper pair wave function can be neglected at temperatures higher than the transition temperature to the {\it p}-wave superfluid. On the other hand, a non-FFLO state with sufficient {\it p}-wave character would not show two peaks (with opposite center-of-mass momenta) in the measured distribution of detected molecules. Rather, it would show a single peak with zero center-of-mass momentum. Thus the proposed detection scheme can be used to rule out such non-FFLO states.



For the sake of simplicity we have focused on one-dimensional situations. The analytical results presented here are however not restricted to 1D systems. In particular, the generalization to higher dimensions of the connection between polarization and triplet-pair correlations (see Eq. (\ref{eqnTripletSingletPolPWF}) is straightforward, although the geometrical configuration of the {\it s-}wave gap profile can be considerably more complex.

In summary, we have shown that there exists a very general and close relation between triplet component and polarization, which explains the findings of \cite{zheng2010,dutta2008}. Our analysis also shows why even a small interaction in the {\it p-}wave channel (with antiparallel spins) can have strong effect on the the stability of the FFLO phase. When the FFLO state is formed, its energy will be lowered by the attractive {\it p-}wave interaction component. This triplet component condensation energy provides a common and simple explanation of the phenomena mentioned in \cite{matsuo1994,shimahara2000,shimahara2002,samokhin2006}. Similarly, a repulsive  {\it p-}wave interaction should reduce the stability of the FFLO phase, as suggested in some of the aforecited works.

We thank R. Hulet and A. Dalgarno for helpful discussions. This work has been supported by MICINN (Spain) through grants FIS2007-65723 and FIS2010-21372, Comunidad de Madrid through MICROSERES grant, Army Research Office with funding from the DARPA OLE program, Harvard-MIT CUA, NSF Grant No. DMR-07-05472, AFOSR Quantum Simulation MURI, AFOSR MURI on Ultracold Molecules, and the ARO-MURI on Atomtronics. One of us (IZ) acknowledges support from Real Colegio Complutense at Harvard.


\begin{thebibliography}{99}
\bibitem{huletPC} A related detection scheme to detect such triplet correlations is the use of photo-association, R. Hulet, Private communication.
\bibitem{note1}  Note that, by construction, the triplet component of the gap is zero, because we have neglected the triplet part of the interaction. This does not preclude however the emergence of triplet pair correlations in the Cooper pair wave function.
\end{thebibliography}
\end{document}